\documentclass[12pt,letterpaper]{article}
\usepackage[a4paper, total={7in, 10in}]{geometry}

\usepackage{graphicx}
\usepackage{helvet}
\usepackage{authblk}
\usepackage{hyperref}
\usepackage{amsmath} 
\usepackage{amssymb} 
\usepackage{orcidlink} 
\usepackage[super,comma,sort&compress]  
   {natbib}

\usepackage[utf8]{inputenc} 
\usepackage[T1]{fontenc}    
\usepackage{url}            
\usepackage{booktabs}       
\usepackage{amsfonts}       
\usepackage{nicefrac}       
\usepackage{microtype}      
\usepackage{xcolor}         
\usepackage{multirow}
\usepackage{threeparttable}
\usepackage{amsmath}
\usepackage{subcaption}
\usepackage{makecell}
\usepackage{placeins}

\makeatletter
\renewcommand{\maketitle}{\bgroup\setlength{\parindent}{0pt}
\begin{flushleft}
  \textbf{\@title}
  
  \@author
\end{flushleft}\egroup}
\makeatother

\title{ECGomics: An Open Platform for AI-ECG Digital Biomarker Discovery}
\date{}

\author[1,\#, \orcidlink{0000-0002-4041-3083}]{Deyun Zhang}
\author[2,\#]{Jun Li}
\author[1]{Shijia Geng}
\author[1]{Yue Wang}
\author[1]{Shijie Chen}
\author[3]{Sumei Fan}
\author[4]{Qinghao Zhao}
\author[2,5,6,*,\orcidlink{0000-0001-7521-5127}]{Shenda Hong}

\affil[1]{HeartVoice Medical Technology, Hefei 230088, China}
\affil[2]{National Institute of Health Data Science, Peking University, Beijing, China}
\affil[3]{College of Integrative Chinese and Western Medicine, Anhui University of Chinese Medicine, Hefei 230012, China}
\affil[4]{Department of Cardiology, Peking University People’s Hospital, Beijing 100044, China}
\affil[5]{State Key Laboratory of Vascular Homeostasis and Remodeling, NHC Key Laboratory of Cardiovascular Molecular Biology and Regulatory Peptides, Peking University, Beijing, China}
\affil[6]{Institute of Medical Technology, Health Science Center of Peking University, Beijing, China}

\affil[$\#$]{These authors contributed equally}
\affil[*]{Correspondence: hongshenda@pku.edu.cn}

\begin{document}

\maketitle

\section*{ABSTRACT}

\textbf{Background:} Conventional electrocardiogram (ECG) analysis faces a persistent dichotomy: expert-driven features ensure interpretability but lack sensitivity to latent patterns, while deep learning offers high accuracy but functions as a black box with high data dependency. We introduce ECGomics, a systematic paradigm and open-source platform for the multidimensional deconstruction of cardiac signals into digital biomarker. \textbf{Methods:} Inspired by the taxonomic rigor of genomics, ECGomics deconstructs cardiac activity across four dimensions: Structural, Intensity, Functional, and Comparative. This taxonomy synergizes expert-defined morphological rules with data-driven latent representations, effectively bridging the gap between handcrafted features and deep learning embeddings. \textbf{Results:} We operationalized this framework into a scalable ecosystem consisting of a web-based research platform and a mobile-integrated solution (\url{https://github.com/PKUDigitalHealth/ECGomics}). The web platform facilitates high-throughput analysis via precision parameter configuration, high-fidelity data ingestion, and 12-lead visualization, allowing for the systematic extraction of biomarkers across the four ECGomics dimensions. Complementarily, the mobile interface, integrated with portable sensors and a cloud-based engine, enables real-time signal acquisition and near-instantaneous delivery of structured diagnostic reports. This dual-interface architecture successfully transitions ECGomics from theoretical discovery to decentralized, real-world health management, ensuring professional-grade monitoring in diverse clinical and home-based settings. \textbf{Conclusion:} ECGomics harmonizes diagnostic precision, interpretability, and data efficiency. By providing a deployable software ecosystem, this paradigm establishes a robust foundation for digital biomarker discovery and personalized cardiovascular medicine.

\section{Introduction}

Electrocardiography (ECG), as a noninvasive, low-cost, and readily accessible physiological signal, has long been regarded as one of the gold standards for cardiovascular disease screening and diagnosis\cite{siontis2021artificial,yang2025ecg}. By capturing patterns of cardiac electrical activity, ECG effectively reveals pathological manifestations such as arrhythmias and myocardial ischemia\cite{li2025electrocardiogram}. However, conventional ECG analysis relies heavily on cardiologists’ visual inspection and experience-based judgment (expert rules). This paradigm, centered on explicit features (e.g., ST-segment depression and T-wave inversion), is inherently limited in its ability to capture subtle fluctuations and high-dimensional nonlinear information embedded in ECG signals, leaving a vast amount of latent data value largely unexplored.

With the advent of the big data era in biomedicine, the emergence of omics research paradigms—such as genomics, transcriptomics, and proteomics—has driven transformative advances\cite{bernatchez2024genomics, guo2025mass}. Rather than focusing on single indicators, these approaches adopt a systems biology perspective to interrogate complex biological systems across multiple dimensions and scales, uncovering hidden patterns and mechanisms\cite{schaffer2021mapping,svinin2025computational,rappaport2026early}. This evolution raises a fundamental question: can ECG signals be conceptualized as a high-dimensional biological data stream and re-examined through an omics-oriented lens to unlock their full life-cycle information content?

In recent years, rapid advances in artificial intelligence (AI), particularly deep learning, have validated the feasibility of this perspective\cite{tao2025multi, zhao2024deep,li2022surveillance,zhang2024integrating}. Beyond surpassing human experts in recognizing conventional cardiac pathologies, AI has achieved breakthroughs in identifying generalized ECG features\cite{noseworthy2022artificial,fiorina2025near}. Accumulating evidence demonstrates that deep neural networks can accurately infer biological attributes such as age, sex, and body mass index from ECG signals\cite{attia2019age,al2025deep, zhou2025continuous}, and can effectively screen for non-cardiovascular conditions, including hyperkalemia, anemia, diabetes, and renal dysfunction\cite{wang2022association,lin2022point, tao2024artificial}. These findings suggest that the heart, as the central driver of systemic circulation and a key target of neurohumoral regulation, produces electrophysiological signals that constitute a holographic projection of whole-body health\cite{ziegler2025neural, bao2025cardiovascular, , ritchie2020basic}. Such latent features are intrinsically linked to an individual’s genetic background, metabolic state, and phenotypic characteristics, paralleling the way genomics deciphers the fundamental codes of life.

Against this backdrop, we formally propose the novel concept of ECGomics. ECGomics represents an interdisciplinary framework integrating signal processing, deep learning, and systems biology. Its objective is to extract high-throughput, automated, and multidimensional features from ECG signals and to systematically elucidate their complex associations with cardiovascular diseases, systemic disorders, genetic backgrounds, and clinical phenotypes. By elevating ECG from a single diagnostic signal to a comprehensive data ensemble rich in biological biomarkers, ECGomics enables a paradigm shift from visual interpretation to data-driven discovery. We anticipate that the establishment of ECGomics will provide a coherent theoretical foundation for unlocking the deep clinical value of ECG data and will accelerate the transition of cardiovascular medicine toward greater precision and intelligence\cite{wang2022clinician}.

\section{Design and development}

\begin{figure}[htbp]
	\centering
    \includegraphics[width=\linewidth]{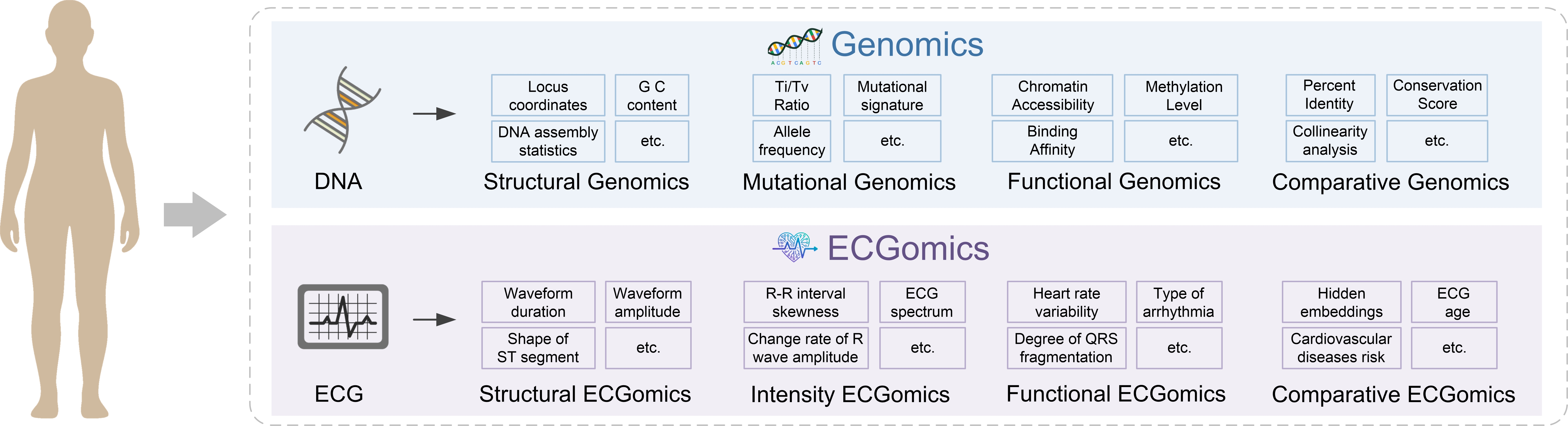}
    \caption{The taxonomic parallel of ECGomics compared with Genomics. The framework establishes a systematic analogy between Genomic and ECGomics to redefine ECG as a high-throughput omics resource. (Top) Genomics deconstructs DNA information into Structural Genomics, Mutational Genomics, Functional Genomics, and Comparative Genomics. (Bottom) Correspondingly, ECGomics deconstructs cardiac signals into four dimensions: Structural ECGomics, Intensity ECGomics, Functional ECGomics, and Comparative ECGomics. This multi-dimensional mapping enables the translation of raw electrical signals into a digital biomarker for systemic health assessment and disease trajectory prediction.} 
    \label{fig:overview}
\end{figure}

\begin{figure}[ht]
	\centering
    \includegraphics[width=\linewidth]{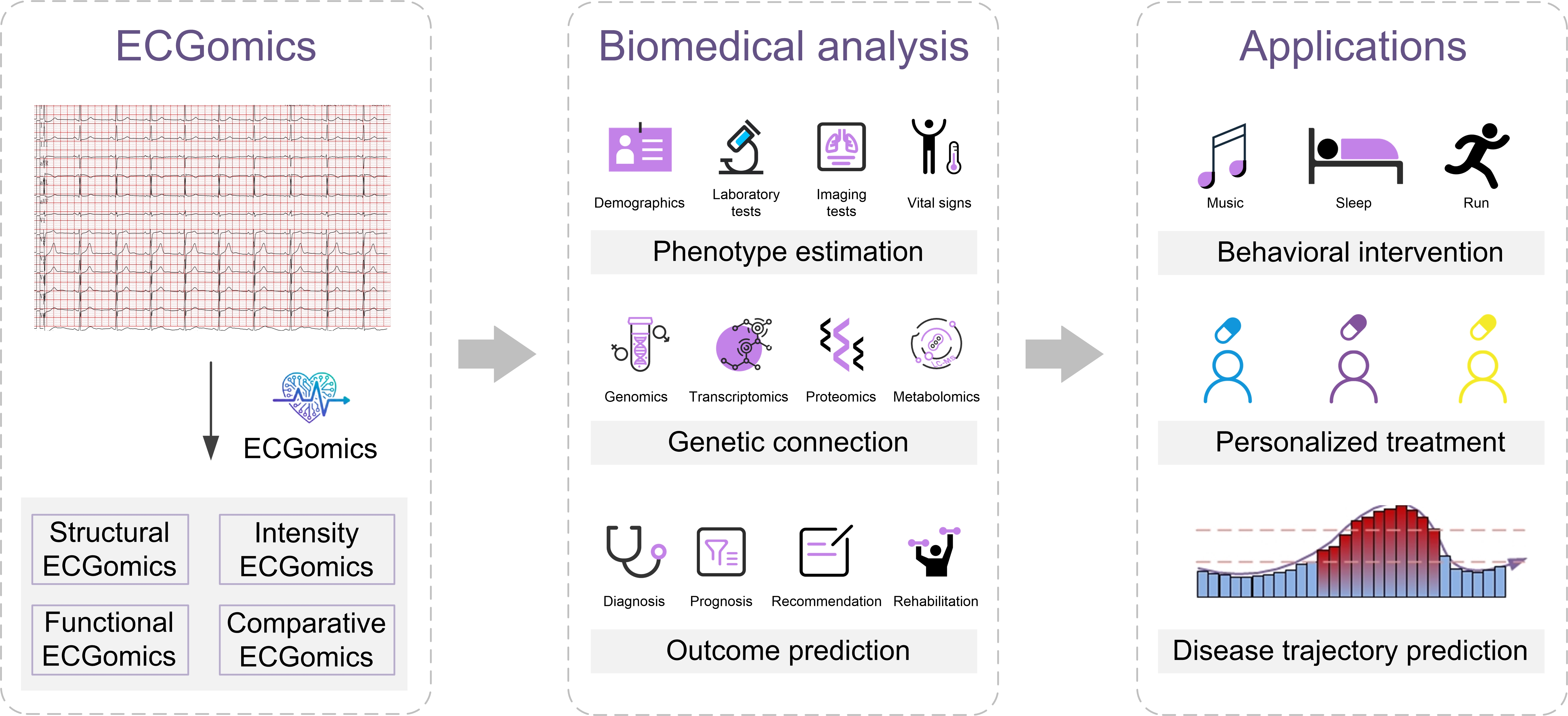}
    \caption{The ECGomics-driven workflow: from digital biomarker to clinical translation. This schematic outlines the end-to-end architecture of the ECGomics paradigm, positioning it as a pivotal nexus that connects cardiovascular phenomics with systemic biology. The integrative framework is organized into three hierarchical layers. ECGomics Data Layer (Left): This layer facilitates high-throughput, multi-modal feature extraction from raw ECG signals. By employing structural, intensity, functional, and comparative ECGomics analyses, it deconstructs complex waveforms into structured digital biomarkers. Biomedical Analysis Layer (Middle): Serving as an integrative hub, this layer correlates ECG-derived signatures with demographics, laboratory biomarkers, multi-modal imaging, and vital signs. It enables deep phenotyping and multi-omics association studies (including genomics, transcriptomics, proteomics, and metabolomics) to support predictive tasks, such as diagnostic classification, prognostic stratification, and clinical decision support. Application Layer (Right): This stage translates high-dimensional insights into actionable clinical pathways, encompassing precision behavioral interventions (e.g., sleep and exercise management), personalized therapeutics, and the dynamic modeling of disease trajectories. Ultimately, this framework exemplifies an integrative paradigm that bridges the gap between cardiovascular phenotypic data and multi-omics information, facilitating advanced precision health management.} 
    \label{fig:stat}
\end{figure}

The concept of ECGomics, as proposed in this study, represents a paradigm shift in cardiovascular medicine, elevating the ECG from a traditional diagnostic tool to a high-throughput, multi-dimensional digital biomarker. This framework is built upon a taxonomic structure that parallels genomics, aiming to transform raw cardiac electrical signals into structured data representations that capture complex temporal and morphological patterns (Fig \ref{fig:overview}). By systematically integrating signal processing with advanced deep learning foundation models, ECGomics establishes a standardized analytical pipeline that bridges the gap between raw physiological data and actionable clinical insights (Fig \ref{fig:stat}).

\subsection{ECGomics definition: four dimensions of ECGomics}

As illustrated in the taxonomic parallel with genomics, ECGomics deconstructs cardiac activity into four distinct yet interconnected dimensions: Structural, Intensity, Functional, and Comparative ECGomics (Fig \ref{fig:overview}). This taxonomy leverages both expert-defined rules and data-driven representations.  

\textbf{Structural ECGomics:} Structural ECGomics corresponds to the morphological characteristics traditionally observed by clinicians, such as P-wave and QRS complex durations and amplitudes. These are extracted through expert-driven frameworks, which serve as the cornerstone for analyzing the physical assembly of the cardiac cycle. 
 
\textbf{Intensity ECGomics:} Intensity ECGomics focuses on the intrinsic physical properties and non-linear dynamics of the signal, quantifying energy distribution and complexity through features such as the ECG spectrum and R-R interval skewness.

\textbf{Functional ECGomics:} Functional ECGomics characterizes the heart's physiological performance and autonomic regulation, utilizing markers like heart rate variability (HRV) and the degree of QRS fragmentation to assess conduction integrity.

\textbf{Comparative ECGomics:} Comparative ECGomics leverages AI-enabled deep representations and association features to benchmark individual signals against vast populations, facilitating the prediction of biological "heart age" and individual traits.

The conceptualization of ECGomics is rooted in our extensive prior research and a robust ecosystem of specialized computational tools. Specifically, we have developed several state-of-the-art deep neural networks based on the Net1D architecture\cite{hong2020holmes}, which were pre-trained on diverse, large-scale datasets to ensure high-dimensional feature fidelity. This foundational work directly fuels the four dimensions of the ECGomics framework: Structural, Intensity, and Functional ECGomics are primarily derived using the ENCASE \cite{hong2017encase} and FeatureDB \cite{fan2025detecting} pipelines, which enable the high-throughput extraction of clinically interpretable morphological and physiological features. In contrast, Comparative ECGomics represents a higher-order synthesis, integrating the predictive power of CardioLearn \cite{hong2020learn} and the vast representation space of ECGFounder \cite{li2025electrocardiogram}. Together, these prior milestones—ranging from expert-driven feature engineering to self-supervised foundation models—provide the multi-layered digital biomarker necessary to map the complex relationships between the electrocardiogram and the systemic health landscape.

\subsection{AI-ECG digital biomarker: perspective of digital biomarkers}

\begin{figure}[htbp]
	\centering
    \includegraphics[width=0.5\linewidth]{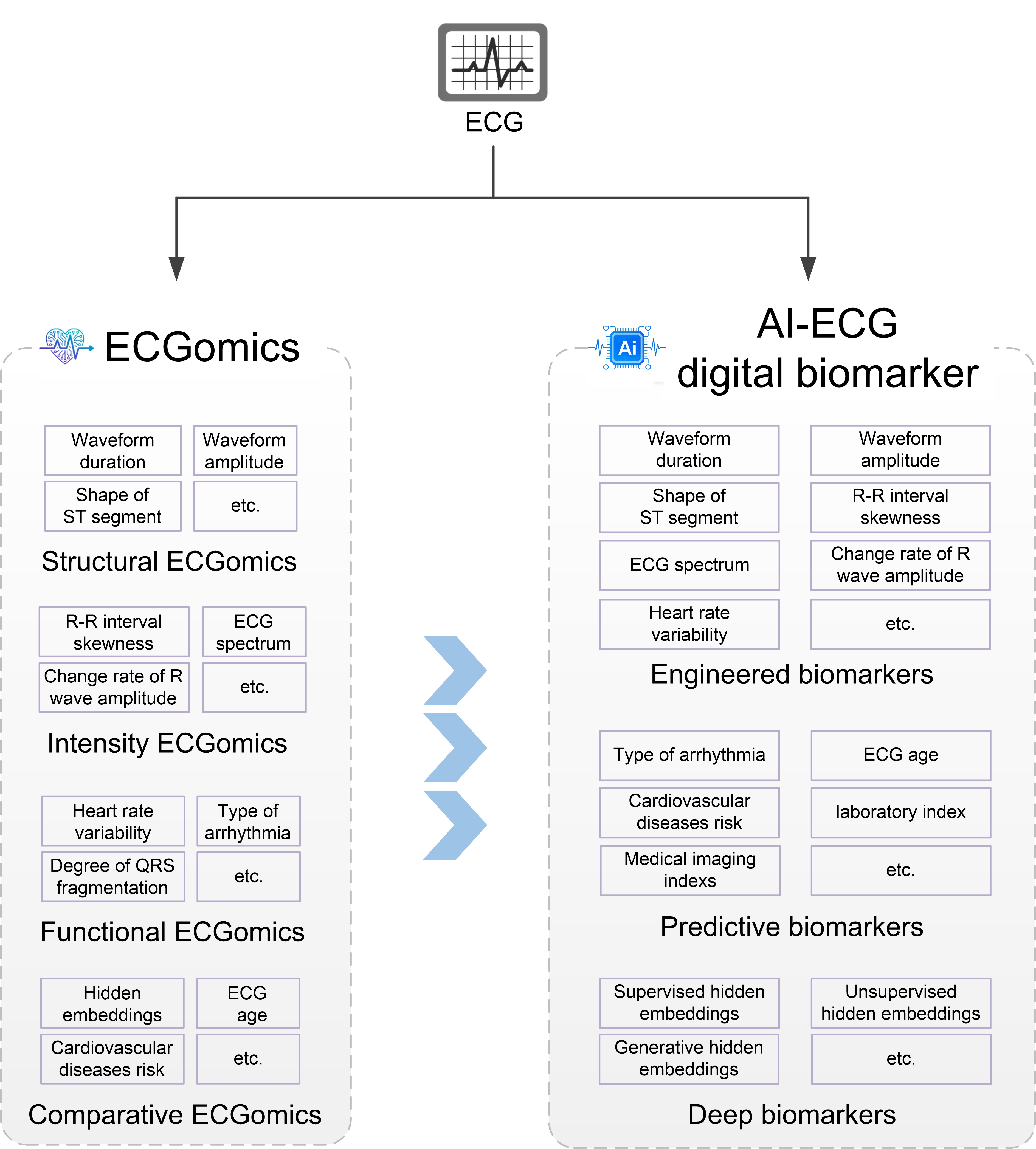}
    \caption{AI-ECG Digital Biomarker overview. Illustrates the classification logic from the underlying raw features to the high-level predictive model, mainly including: engineered biomarkers (corresponding to structural, intensity and functional dimensions, such as waveform duration, spectrum and heart rate variability), predictive biomarkers (corresponding to comparison dimensions, such as cardiac age and disease risk prediction), and deep biomarkers based on deep learning (such as various hidden layer embedded features).} 
    \label{fig:2-1}
\end{figure}

Fig \ref{fig:2-1} delineates a hierarchical taxonomy of AI-ECG digital biomarkers, categorizing them into three progressive layers that align with the ECGomics framework, comprising Structural, Intensity, Functional, and Comparative dimensions. This systematic deconstruction bridges traditional clinical interpretation with advanced computational phenotyping, providing a comprehensive understanding of disease.

At the foundational level, engineered biomarkers encompass the Structural, Intensity, and Functional dimensions of ECGomics. These are derived through expert-defined rules and include morphological parameters such as waveform duration and amplitude, which quantify the physical assembly of the cardiac cycle. Concurrently, these markers capture non-linear dynamics and autonomic regulation through metrics like the ECG spectrum, R-R interval skewness, and heart rate variability (HRV), thereby providing a multi-dimensional assessment of cardiac conduction integrity and energy distribution.

The middle tier consists of predictive biomarkers, which primarily align with the Comparative ECGomics dimension. By leveraging AI-enabled association features, these biomarkers benchmark individual electrophysiological signals against vast population datasets. This comparative analysis facilitates high-level clinical inferences, including the determination of "ECG age," the stratification of cardiovascular disease risk, and the cross-modal prediction of laboratory indices and medical imaging features that transcend traditional visual analysis.

Finally, the architecture culminates in deep biomarkers, representing the pinnacle of data-driven representation. These biomarkers consist of hidden embeddings—latent features extracted via supervised, unsupervised, and generative deep learning architectures. Unlike engineered features, deep biomarkers encapsulate complex, high-dimensional patterns within the cardiac signal that are often imperceptible to the human eye, offering a granular, digital representation of the heart's underlying physiological state.

\subsection{Analytical workflow: from extraction to application}

The conceptualization and practical execution of the ECGomics framework are operationalized through a rigorous three-stage analytical pipeline consisting of multi-dimensional extraction, systematic biomedical mapping, and clinical translation applications. 

In the initial extraction stage, we implement a sophisticated hybrid strategy that harmonizes clinical interpretability with high-dimensional data expression. This process utilizes our proprietary FeatureDB framework to extract a structured vector of expert-defined morphological and physiological metrics, such as traditional wave intervals and amplitudes, while simultaneously employing the ECGFounder foundation model to derive latent digital biomarkers. By processing raw signals through the Net1D architecture, we capture high-dimensional hidden-layer embeddings that encompass subtle, subclinical signal perturbations often invisible to human observers. This integration of a glass-box expert engine and a deep representation engine ensures that the resulting ECGomics serves as a comprehensive digital biomarker of cardiac and systemic health.

Building upon this data foundation, the biomedical analysis stage serves as a conceptual bridge that maps ECGomics features onto a broad landscape of clinical and biological information, closely paralleling the integrative logic of multi-omics research. This phase involves deep phenotyping through statistical correlation and integrative modeling to link ECGomics vectors with diverse clinical layers, ranging from laboratory biomarkers, such as NT-proBNP, to cardiac imaging metrics, including left ventricular ejection fraction. Furthermore, this stage facilitates the identification of electro-genetic signatures by correlating deep representations with genomic data, thereby revealing the underlying genetic determinants of cardiac electrical patterns. Unlike traditional electrocardiography, which focuses predominantly on localized cardiac pathology, the ECGomics paradigm expands into systemic mapping, positioning the heart as a holistic sensor capable of reflecting metabolic, renal, and endocrine dysfunctions. By training on the concatenated vector of expert and deep features, the model performs personalized risk stratification and disease detection. 

The final application stage translates these systemic insights into actionable clinical decisions through robust predictive modeling and personalized medicine strategies. By utilizing advanced machine learning algorithms, notably XGBoost, the framework effectively handles the heterogeneous and high-dimensional nature of the fused feature sets to identify and forecast diverse clinical outcomes. This modeling paradigm enables precise and personalized risk stratification across a wide spectrum of conditions, ranging from acute cardiovascular events to non-cardiovascular systemic diseases, such as chronic kidney disease. Ultimately, the ECGomics pipeline facilitates a transition toward proactive, precision care, supporting the identification of novel digital biomarkers for non-cardiac diseases' opportunistic screening (e.g., CKD, anemia), the design of individualized behavioral interventions, and the high-resolution prediction of long-term disease trajectories for global patient management.




\subsection{Online tool development: ECGomics analysis platform}

\begin{figure}[htp]
	\centering
    \includegraphics[width=\linewidth]{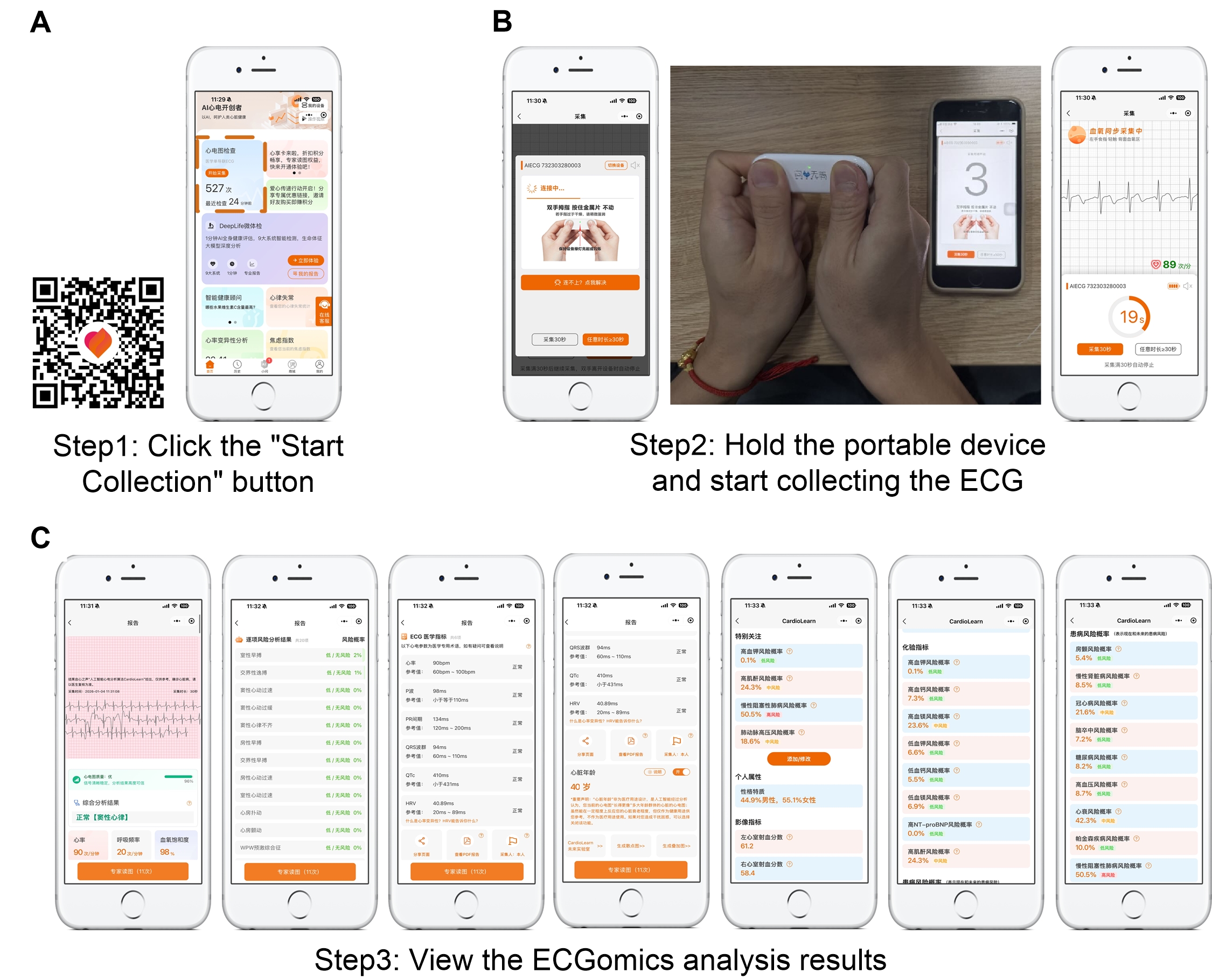}
    \caption{Operation Process for Portable ECG Collection and ECGomics Analysis. (A) Navigate to the primary mobile application dashboard and execute the "Start Collection" command to initialize the automated data acquisition and processing sequence. (B) Securely hold the handheld portable ECG transducer with both hands to establish a stable connection; the interface provides real-time, synchronous visual feedback of the collection progress (e.g., the countdown and live waveform), ensuring the integrity of the captured electrocardiographic signals. (C) Upon successful completion of the recording, the platform performs near-instantaneous processing to deliver a comprehensive analysis report. This output facilitates a granular assessment of cardiac health by integrating diverse interpretive dimensions, including high-fidelity waveform visualization, automated rhythm diagnostics (e.g., Heart Rate, PR interval, QRS duration), and advanced predictive metrics such as biological heart age and cardiovascular risk stratification.} 
    \label{fig:5}
\end{figure}

To facilitate the broad adoption of the proposed framework and bridge the gap between theoretical research and clinical utility, we have developed a dedicated web-based (Figure \ref{fig:4}) and phone-based (Figure \ref{fig:5}) platform titled ECGomics Analysis, accessible at \url{https://github.com/PKUDigitalHealth/ECGomics}. This online tool serves as a practical, open-access gateway for the global research community, allowing for the high-throughput implementation of the ECGomics pipeline without requiring local high-performance computing clusters or deep learning expertise. The platform is engineered to automate the transition from raw physiological signals to structured digital phenotypes by integrating our core backend engines, including the Net1D-based FeatureDB protocol for expert-driven morphological characterization and the ECGFounder foundation model for latent representation mining. Through a user-friendly interface, researchers can upload raw ECG recordings and receive a comprehensive deconstruction of the signal across the structural, intensity, functional, and comparative dimensions.

A primary objective of this development is to ensure data transparency and foster collaborative innovation in the field of precision medicine. Upon completion of the automated analysis, the platform allows users to download the full suite of extracted ECGomics features as structured datasets. These downloadable resources include both clinically interpretable metrics and high-dimensional hidden-layer embeddings, enabling users to perform independent downstream investigations such as large-scale phenotype association studies or the discovery of novel electro-genetic biomarkers. By providing this standardized tool for feature extraction and digital biomarker identification, the platform democratizes access to advanced AI-enabled ECG analysis, supporting the identification of systemic health indicators and the high-resolution prediction of disease trajectories across diverse patient populations.

\begin{figure}[ht]
	\centering
    \includegraphics[width=\linewidth]{fig/Figs4.jpg}
    \caption{The user interface of the ECGomics platform and usage steps are displayed on the right side. (A) Allows for the configuration of signal acquisition settings, including SampleRate (Hz), adcGain, and adcZero (mV), ensuring the raw data is correctly calibrated for subsequent processing. (B) Select Sample Data from a dropdown menu or upload custom .npy files for analysis. (C) Shows the raw numerical representation of the ECG signal, allowing for immediate verification of the digital data string before visualization. (D) Provides a high-fidelity visualization of the 12-lead ECG waveform. It includes a standard grid for clinical assessment and an Export function for extracting digitized waveform data. (E) Positioned at the bottom of the left panel, this Generate ECGomics control serves as the execution command to initiate the AI-driven biomarker extraction process. (F) A utility feature that allows users to switch the interface language between Chinese and English, facilitating international research collaboration. (G) Represents the core output of the platform, categorized into the four distinct dimensions of the ECGomics taxonomy. All the analysis results can be exported in CSV format by clicking the green icon on the right.} 
    \label{fig:4}
\end{figure}

\section{Validation of the ECGomics framework in application}

The ECGomics framework has been rigorously applied to a series of representative scenarios, demonstrating its versatility and robustness as a multi-dimensional analytical paradigm for both cardiac-specific and systemic health assessment.

\subsection{Atrial fibrillation detection}

In our previous work, we developed ENCASE\cite{hong2019combining}, a two-stage analytical method that epitomizes the integration of expert-driven morphological characterization and data-driven deep representations. By utilizing short, single-lead ECG recordings from the PhysioNet 2017 Challenge, this approach achieved a superior F1 score of 0.825 across diverse rhythm categories (Normal, AF, and Other). A critical finding of this study was through feature attribution analysis, which revealed that expert-defined morphological features and deep neural representations provide complementary physiological information. This synergy allowed the model to maintain high diagnostic accuracy even in noisy or brief signal segments, establishing the feasibility of the ECGomics framework for reliable, real-time arrhythmia screening.

\subsection{Predicting atrial fibrillation recurrence after cryoablation}

To explore the prognostic potential of the framework, we investigated the prediction of atrial fibrillation recurrence following cryoablation therapy \cite{song2024prediction}. By analyzing standard 12-lead ECGs from a cohort of 201 patients, we constructed predictive models that integrated baseline AF subtypes with digital biomarker—the high-dimensional differences between pre- and post-procedural deep representations. The optimized XGBoost model achieved an AUC of 0.872 and an overall accuracy of 0.902, significantly outperforming traditional clinical risk scores. These results underscore the unique ability of ECGomics to extract biomarkers from subtle shifts in the cardiac electrical activity, providing a quantitative basis for individualized risk assessment and the optimization of post-ablation follow-up strategies.

\subsection{Detection of severe coronary stenosis in apparently normal ECGs}

The ECGomics framework has also demonstrated remarkable efficacy in uncovering hidden cardiovascular risks in apparently normal ECGs \cite{xue2024screening}. In a study involving 392 patients, we applied deep transfer learning to identify individuals with severe coronary stenosis who displayed no overt abnormalities under conventional clinical inspection. While standard ECG interpretation showed limited sensitivity (0.545), the integration of deep ECGomics features with multi-modal clinical variables improved the sensitivity to 0.848 with an AUC of 0.847. This application highlights the framework's strength in Comparative ECGomics, where the model learns to detect subtle subclinical perturbations that escape human observation, effectively serving as a non-invasive screening tool for high-risk coronary artery disease.

\subsection{Maternal health management based on ECGomics}

Extending the reach of ECGomics to specialized populations, we evaluated the reliability and validity of a novel AI-enabled portable device, WenXinWuYang, for cardiac monitoring in pregnant women\cite{wang2025reliability}. Pregnancy introduces significant physiological changes that increase the cardiac load, yet continuous monitoring is often hindered by the inconvenience of frequent hospital-based 12-lead ECGs. In a comparative study of 99 pregnant women across all trimesters, the ECGomics-based system demonstrated high diagnostic consistency (>0.900) and strong correlation with clinical gold standards for heart rate ($r=0.957$) and QT intervals ($r=0.774$). The system achieved a sensitivity of 0.842 and a specificity of 0.975 in detecting arrhythmias, proving that the integration of deep learning-based representation with portable hardware can provide professional-grade monitoring in real-world maternal care. This application confirms that ECGomics can effectively facilitate the transition toward proactive, home-based health management for vulnerable populations.

\subsection{Summary of Above Studies}

Collectively, these case studies validate the wide-ranging utility of the ECGomics paradigm across diagnostic, prognostic, and risk-stratification tasks. Three core strengths are consistently evidenced: first, the synergistic integration of morphological, intensity, and deep representations significantly enhances predictive power over traditional methods; second, the use of interpretability tools (e.g., SHAP) bridges the gap between AI-driven discovery and clinical trust; and third, the framework possesses a unique capacity to reveal hidden systemic and cardiovascular markers that are otherwise imperceptible. Together, these milestones reinforce ECGomics as a transformative, systematic paradigm for the future of precision cardiovascular medicine.

\section{Comparison of ECG predictive workflows}

\begin{figure}[ht]
	\centering
    \includegraphics[width=1\linewidth]{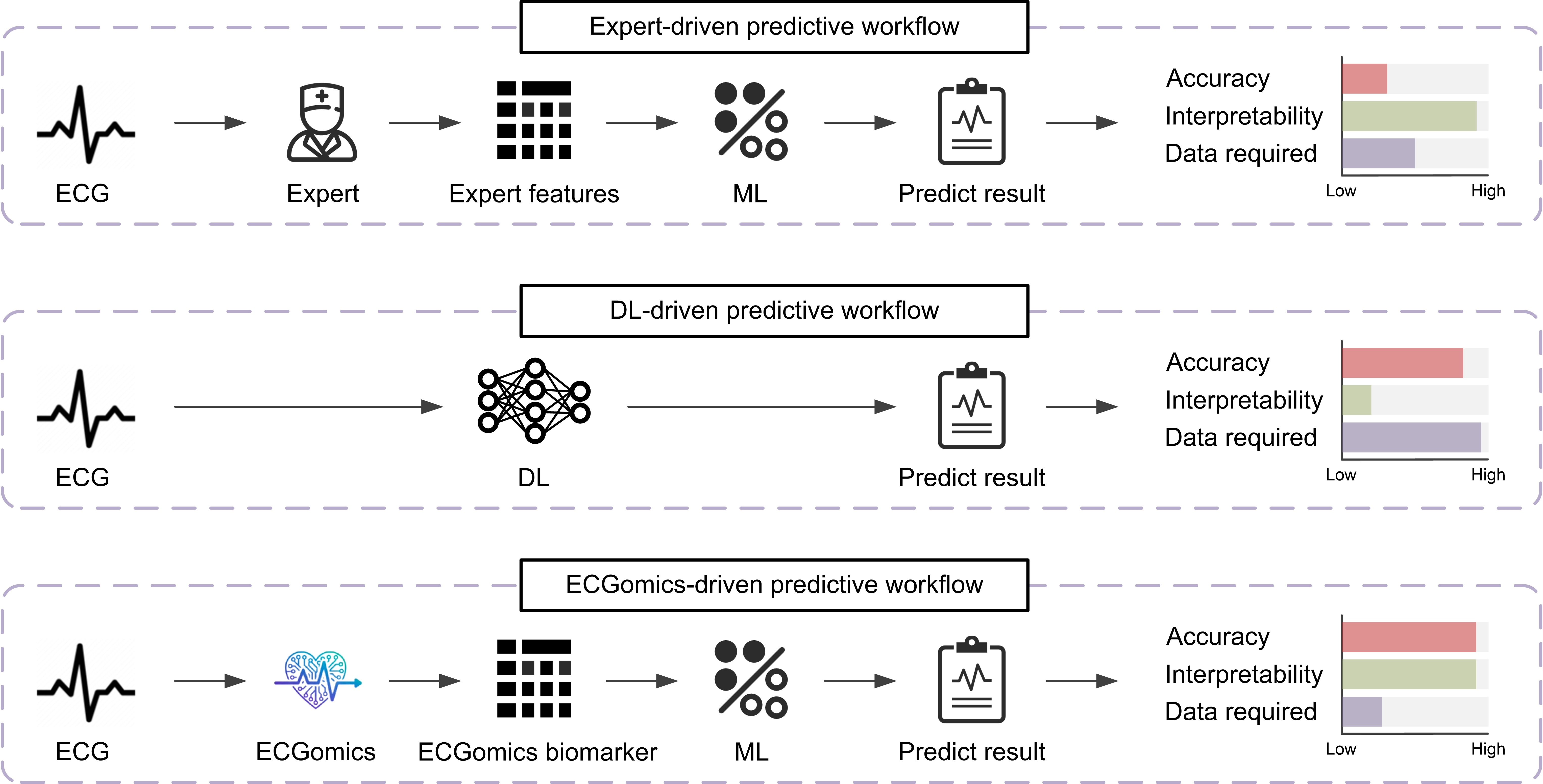}
    \caption{Comparison of different ECG prediction workflows. This figure illustrates the methodological evolution and performance trade-offs across three ECG analysis paradigms. (Top) The Expert feature-driven workflow utilizes handcrafted features based on clinical knowledge, offering high interpretability but limited accuracy due to the omission of latent signal patterns. (Middle) The DL-driven workflow achieves superior accuracy through end-to-end feature learning but functions as a "black box" with low interpretability and high data dependency. (Bottom) The proposed ECGomics-driven workflow synergizes expert-defined morphology with deep latent representations. As shown in the bar charts (right), the ECGomics approach achieves a high balance between accuracy and interpretability while maintaining a low data requirement, demonstrating its superior utility for precision medicine applications.} 
    \label{fig:pred}
\end{figure}

As illustrated in Fig \ref{fig:pred}, we compared three distinct ECG predictive workflows to highlight the unique advantages of the ECGomics-driven approach in balancing accuracy, interpretability and data required.

\textbf{Expert-driven Workflow:} This traditional approach relies on manually defined morphological features extracted by clinical experts. While it offers the highest level of interpretability due to its grounding in established physiological rules, it suffers from limited accuracy as it fails to capture the high-dimensional latent information within the signal. However, it remains highly efficient in terms of data requirement, necessitating only a small, well-annotated dataset for model training.

\textbf{Deep Learning (DL)-driven Workflow:} In contrast, the end-to-end DL approach utilizes raw ECG signals to achieve superior accuracy by automatically learning complex patterns. However, this gain in performance comes at the cost of being a "black box," leading to very low interpretability. Furthermore, it is highly data-intensive, requiring massive amounts of labeled data to reach its full potential.

\textbf{ECGomics-driven Workflow:} The proposed ECGomics paradigm serves as a synergistic middle ground. By integrating the glass-box interpretability of expert features with the powerful expressive capacity of deep latent representations (via models like ECGFounder), it achieves an accuracy comparable to DL-driven methods while significantly outperforming them in interpretability. Crucially, by leveraging pre-trained foundation models, the ECGomics workflow maintains a relatively low data requirement for downstream tasks, making it a highly practical and robust solution for personalized medicine.

\section{Using instruction}

To facilitate the broad adoption of the proposed framework and bridge the gap between theoretical innovation and clinical utility, the ECGomics paradigm is operationalized through a web-based ecosystem (Figure \ref{fig:4}) and a phone-based ecosystem (Figure \ref{fig:5}), ECGomics Analysis, available at \url{https://github.com/PKUDigitalHealth/ECGomics}. This interactive platform serves as a practical implementation of the ECGomics pipeline, providing a high-throughput interface for the intuitive exploration and real-time visualization of multi-dimensional cardiac signatures. By lowering the technical barriers associated with complex signal processing and deep learning, the platform enables researchers and clinicians to upload raw ECG recordings—or use preloaded benchmark examples—to systematically deconstruct the cardiac signal into a comprehensive and structured feature space.

Furthermore, to address scenarios where standard ECG signals are difficult to obtain, we have developed a mobile ECGomics platform compatible with portable single-lead devices. This platform, integrated within the WeChat Mini-Program ecosystem, leverages portable sensors for signal acquisition and is capable of performing rapid, automated extraction and delivery of AI-ECG digital biomarkers in near real-time.\\

\textbf{\large{Option 1: Use web-based ECGomics platform}}

The following steps define the end-to-end functional chain of the ECGomics platform, a specialized system for the systematic discovery and extraction of AI-ECG digital biomarkers.

\textbf{Step 1: Configuration of the precision parameter}

This foundational stage ensures that the platform accurately interprets the physical properties of the electrocardiographic signal. Users configure critical acquisition parameters, including the Sampling Rate (defining temporal resolution), ADC Gain (normalizing signal amplitude), and Zero-point Voltage (establishing the baseline calibration). The alignment between these digital settings and the physical acquisition hardware is paramount to maintain signal fidelity during downstream analysis (Figure \ref{fig:4}A).

\textbf{Step 2: High-fidelity data ingestion via npy format}

The platform utilizes the .npy format to facilitate data input, a choice that ensures the efficient storage of multi-dimensional numerical arrays while preserving the raw integrity of the ECG signal. In the data management module, users may either perform a localized upload of their proprietary files via the "Upload .npy" interface or utilize standardized "Sample Data" for benchmark testing (Figure \ref{fig:4}B). Upon submission, the platform executes automated verification of file integrity to establish a robust dataset for the analytical engine.

\textbf{Step 3: Multi-lead visualization and signal verification}

To bridge the gap between raw data and clinical insight, the uploaded signal is rendered into an intuitive 12-lead ECG waveform visualization. This interface allows researchers to perform a preliminary qualitative review of rhythm patterns and morphological characteristics. Furthermore, an "Export ECG data" function is provided, enabling users to archive the digitized waveforms or conduct secondary independent validation (Figure \ref{fig:4}C,D).

\textbf{Step 4: AI-driven execution of the ECGomics engine}

The core of the discovery process is initiated by executing the "Generate ECGomics" command. This triggers the platform's proprietary AI algorithms to perform a multi-dimensional deconstruction of the cardiac signal. The engine extracts digital biomarkers across four interconnected dimensions: Structural ECGomics, Intensity ECGomics, Functional ECGomics, and Comparative ECGomics (Figure \ref{fig:4}E).

\textbf{Step 4: Dimensional result synthesis and research}

The workflow culminates in the presentation of a structured ECGomics report, which categorizes biomarkers into modules such as demographics-related traits (e.g., Heart Age), laboratory test correlations, and disease probability scores. To support further scientific inquiry, each analytical module allows for data extraction in csv format. This ensures seamless compatibility with external statistical software, facilitating high-level research integration and large-scale clinical population studies (Figure \ref{fig:4}F).\\

\textbf{\large{Option 2: Use phone-based ECGomics platform with portable devices}}

\textbf{Step 1: Procedural activation}

Upon launching the mobile health application (WeChat mini program), the user accesses a centralized dashboard that integrates personal health metrics with historical ECG records. The diagnostic sequence is initiated by engaging the prominent "Start Collection" interface. This action triggers the system's initialization phase, transitioning the interface into a preparatory mode that displays essential operational guidelines to ensure procedural compliance before hardware coupling (Figure \ref{fig:5}A).

\textbf{Step 2: Synchronized signal acquisition}

Following initialization, the mobile terminal automatically establishes a secure near-field communication link (typically via Bluetooth) with the portable ECG sensor. Once pairing is confirmed, the interface guides the user to maintain a standardized bimanual posture to ensure optimal electrode-skin conductivity. The device then captures ECG in real-time, synchronizing the raw data stream with the mobile unit, while a dynamic visual indicator (e.g., collection countdown) provides immediate feedback on the progress of acquisition (Figure \ref{fig:5}B).

\textbf{Step 2: Cloud-based ECGomics analysis and result delivery}

Upon completion of signal acquisition, the raw data are securely transmitted to the CardioLearn cloud service platform for processing. The cloud infrastructure deploys professional AI algorithms to execute a multi-dimensional ECGomics analysis on the dataset. The processed results are rapidly returned to the mobile interface as a structured report, encompassing high-fidelity waveform visualization, quantitative rhythm parameters, and functional cardiac assessments, thereby providing the user with immediate, interpretable insights into their cardiac health (Figure \ref{fig:5}C).



\section{Conclusion}

This work establishes ECGomics as a systematic paradigm that integrates morphological, associative, and deep representations to bridge the gap between traditional feature engineering and end-to-end deep learning. Validated through diverse case studies—ranging from arrhythmia detection to identifying occult coronary stenosis—ECGomics demonstrates superior diagnostic accuracy and interpretability while maintaining data efficiency. These capabilities allow the framework to uncover subtle physiological patterns essential for personalized medicine. Despite these strengths, challenges regarding data scale, annotation quality, and the need for rigorous multi-center validation persist. Future advancements must focus on refining model interpretability and integrating ECGomics with multi-omics (e.g., genomics) and multimodal clinical data. Collectively, these developments position ECGomics as a foundational paradigm to advance precision cardiovascular diagnosis, risk stratification, and therapeutic decision-making.

\section*{ACKNOWLEDGMENTS}


This study was supported by the National Natural Science Foundation of China (62102008, 62172018), CCF-Tencent Rhino-Bird Open Research Fund (CCF-Tencent RAGR20250108), CCF-Zhipu Large Model Innovation Fund (CCF-Zhipu202414), PKU-OPPO Fund (BO202301, BO202503), Research Project of Peking University in the State Key Laboratory of Vascular Homeostasis and Remodeling (2025-SKLVHR-YCTS-02), and Fan Sumei scientific research start-up funds (DT2400000509).

\section*{AUTHOR CONTRIBUTIONS}

Concept and design: D.Z., J.L., and S.H.  
Acquisition, analysis, or interpretation of data: S.G., Y.W., S.C., S.F., Q.Z., and S.H.  
Drafting of the manuscript:  D.Z., J.L., and S.H.  Critical revision of the manuscript for important intellectual content: D.Z., J.L., S.G., Y.W., S.C., S.F., Q.Z., and S.H.
Tool development:  D.Z., J.L., S.C., and S.H.  
Obtained funding: S.F. and S.H.  
Supervision: S.H. 



\section*{DECLARATION OF INTERESTS}


The authors declare that they have no competing interests.

\newpage


\bibliography{reference}

\bigskip


\newpage

\end{document}